\documentclass{article}
\usepackage{PRIMEarxiv}

\usepackage[utf8]{inputenc} 
\usepackage[T1]{fontenc}    
\usepackage{hyperref}       
\usepackage{url}            
\usepackage{booktabs}       
\usepackage{amsfonts}       
\usepackage{nicefrac}       
\usepackage{microtype}      
\usepackage{lipsum}
\usepackage{fancyhdr}       
\usepackage{graphicx}       
\graphicspath{{media/}}     
\usepackage{float}

\usepackage{xspace}
\usepackage{amssymb}
\usepackage{booktabs}
\usepackage{multirow}
\usepackage{arydshln}
\usepackage{amssymb}
\usepackage{hhline}
\usepackage{tcolorbox}
\usepackage{verbatim}
\usepackage{listings}
\usepackage{fancyvrb}
\usepackage{xcolor}
\usepackage{tikz}
\usepackage{graphicx}

\newcommand{\swebench}{SWE-bench\xspace}
\newcommand{\swebenchjava}{\texttt{SWE-bench-java-verified}\xspace}

\definecolor{darkgreen}{rgb}{0.0, 0.5, 0.0}

\newcommand{\multiswebench}{Multi-SWE-bench\xspace}
\newcommand{\filledcirclepercent}[1]{
    \begin{tikzpicture}
    \draw (0,0) circle (0.2cm);
    \fill[black] (0,0) -- (0.2cm,0) arc [start angle=0, end angle=#1*3.6, radius=0.2cm] -- cycle;
    \end{tikzpicture}
}

\title{SWE-bench-java: A GitHub Issue Resolving\\Benchmark for Java}

\author{
Daoguang Zan\textsuperscript{1}\thanks{Equal contribution}
~~Zhirong Huang\textsuperscript{1\textasteriskcentered{}}~~Ailun Yu\textsuperscript{2\textasteriskcentered{}} \\ 
\textbf{Shaoxin Lin\textsuperscript{3}}~~\textbf{Yifan Shi\textsuperscript{2}}~~\textbf{Wei Liu\textsuperscript{2}}~~\textbf{Dong Chen\textsuperscript{3}}~~\textbf{Zongshuai Qi\textsuperscript{2}}
\\
\textbf{Hao Yu
\textsuperscript{2}}~~\textbf{Lei Yu
\textsuperscript{1}}~~\textbf{Dezhi Ran\textsuperscript{2}}~~\textbf{Muhan Zeng\textsuperscript{3}}~~\textbf{Bo Shen\textsuperscript{3}}~~\textbf{Pan Bian\textsuperscript{3}}
\\
\textbf{Guangtai Liang\textsuperscript{3}}~~\textbf{Bei Guan\textsuperscript{1}}~~\textbf{Pengjie Huang\textsuperscript{4}}~~\textbf{Tao Xie\textsuperscript{2}}~~\textbf{Yongji Wang\textsuperscript{1}}~~\textbf{Qianxiang Wang\textsuperscript{3}}
\\
\textsuperscript{$^1$}Chinese Academy of Science
\textsuperscript{$^2$}Peking University
\textsuperscript{$^3$}Huawei Co., Ltd.
\textsuperscript{$^4$}Lingzhi-zhiguang Co., Ltd
}

\begin{document}
\maketitle

\vspace{-1cm}
\begin{center}
\raisebox{-0.3\height}{\includegraphics[height=0.5cm]{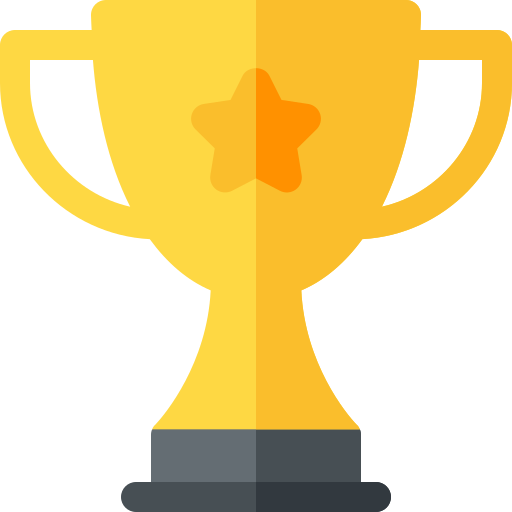}} \hspace{0.1cm} \href{https://multi-swe-bench.github.io}{https://multi-swe-bench.github.io}

\raisebox{-0.3\height}{\includegraphics[height=0.5cm]{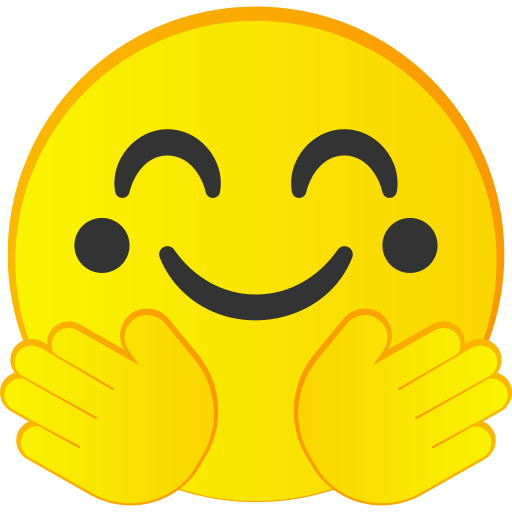}} \hspace{0.1cm} \href{https://huggingface.co/datasets/Daoguang/Multi-SWE-bench}{https://huggingface.co/datasets/Daoguang/Multi-SWE-bench}

\raisebox{-0.3\height}{\includegraphics[height=0.5cm]{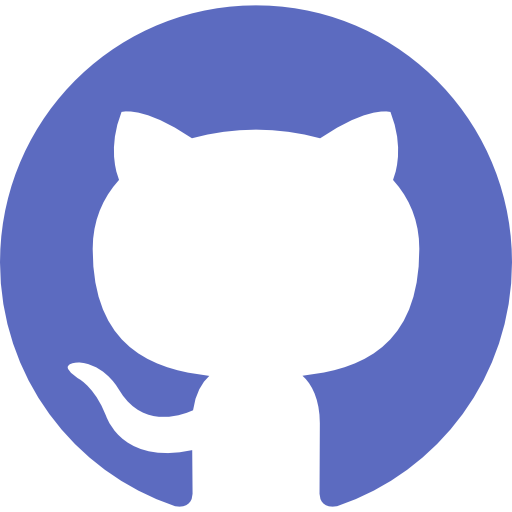}} \hspace{0.1cm} \href{https://github.com/multi-swe-bench/multi-swe-bench-env}{https://github.com/multi-swe-bench/multi-swe-bench-env}
\end{center}
\vspace{1cm}

\begin{abstract}
GitHub issue resolving is a critical task in software engineering, recently gaining significant attention in both industry and academia.
Within this task, SWE-bench~\cite{swe-bench} has been released to evaluate issue resolving capabilities of large language models (LLMs), but has so far only focused on Python version.
However, supporting more programming languages is also important, as there is a strong demand in industry.
As a first step toward multilingual support, we have developed a Java version of SWE-bench, called \swebenchjava.
We have publicly released the dataset, along with the corresponding Docker-based evaluation environment and leaderboard, which will be continuously maintained and updated in the coming months.
To verify the reliability of \swebenchjava, we implement a classic method SWE-agent~\cite{yang2024sweagent} and test several powerful LLMs on it.
As is well known, developing a high-quality multi-lingual benchmark is time-consuming and labor-intensive, so we welcome contributions through pull requests or collaboration to accelerate its iteration and refinement, paving the way for fully automated programming.
\end{abstract}

\section{Introduction}
\label{sec:introduction}

Automating software engineering tasks with large language models (LLMs) has gained considerable attention~\cite{nl2code,zheng2023survey,zhang2023survey}. 
Beyond code generation, the issue resolving task proposed by SWE-bench~\cite{swe-bench} changes the role of LLMs from code assistants to fully autonomous AI programmers.
SWE-bench contains $2,294$ issues from $12$ widely-used open-sourced Python libraries.
LLMs are tasked to generate a patch based on the issue description along with the buggy code repository.
Within less than one year, the resolving rate on SWE-bench lite increased from $0.33\%$~\cite{swe-bench} (for RAG+GPT3.5) to $43.00\%$~\cite{codeStory} (for CodeStory Aide+Mixed Models).
SWE-bench lite is a subset of $300$ issues selected from SWE-bench, chosen for their relatively clear descriptions and simple fix solutions.

SWE-bench~\cite{swe-bench} is centered on Python, which confined its evaluation to LLMs in Python-related fields such as data processing and artificial intelligence.
However, it does not cover other common and necessary fields like web applications, mobile applications, and system programming, which rely on other programming languages~\cite{defects4j,huang2019empirical}.
Therefore, as the first step in moving towards a multi-lingual issue resolving benchmark, 
we choose to develop a Java version of SWE-bench for the following two reasons: 
\begin{enumerate}
    \item \textbf{Popularity.} 
    Java enjoys widespread popularity, making it one of the most widely adopted programming languages in the industry, particularly in fields like finance, cloud services, and Android application development.
    According to the TIOBE index\footnote{\url{https://www.tiobe.com/tiobe-index}} for August 2024, Java ranks among the top 4 languages, following Python, C, C++.
    With an active developer community, Java continues to grow, as evidenced by Oracle's data\footnote{\url{https://blog.jetbrains.com/idea/2024/07/is-java-still-relevant-nowadays}} showing $120$ million developers worldwide, with over $1$ million new developers added annually.
    \item \textbf{Platform Independence.} 
    Java programs run on a virtual machine, which automatically manages memory similar to Python, and compiles to Java byte-code that is interpreted by the virtual machine.
    While C and C++ are preferred for system programming due to their performance and memory efficiency, we believe that language models are not primarily designed to address performance issues. Therefore, we chose Java over the more performance-focused C/C++ as the first step. 
\end{enumerate}

This paper proposes a Java version SWE-bench, named \swebenchjava. We describe the details of dataset construction, the main challenges, and potential problems. With \swebenchjava, we also evaluate the performance of SWE-agent~\cite{yang2024sweagent} with the state-of-the-art models including GPT-4o~\cite{gpt4o}, GPT-4o-mini~\cite{gpt4omini}, DeepSeek-V2~\cite{deepseeklm}, DeepSeekCoder-V2~\cite{deepseekcoderv2}, Doubao-pro~\cite{doubao}.
Overall, the contributions of this paper are as follows:
\begin{itemize}
    \item We meticulously created and manually verified the \swebenchjava benchmark, marking the first step in establishing a multilingual GitHub issue-resolving benchmark with a focus on Java. We plan to support more programming languages and make continuous improvements in the coming months. We also encourage the community to contribute by submitting pull requests to collaborate in advancing this field.
    \item We open-sourced the dataset, along with a comprehensive evaluation Docker environment and a leaderboard, to advance further research in this field.
    \item We implemented SWE-Agent~\cite{yang2024sweagent} on \swebenchjava and derived several insightful findings that enhance our understanding of issue resolving in Java projects.
\end{itemize}

\section{\multiswebench}
\subsection{Benchmark Construction}

\subsubsection{Workflow Overview}
\label{sec:workflow_overview}
We construct \swebenchjava by following the work of SWE-bench~\cite{swe-bench}. Specifically, the workflow of constructing this benchmark comprises five phases:

\begin{enumerate}
    \item \textbf{Candidate repository collection.}
    We collect candidate repositories for \swebenchjava construction from two sources: 
    (1) Popular Java repositories on GitHub, which are collected by requesting a list of Java repositories sorted by their stars via GitHub API\footnote{\url{https://docs.github.com/en/rest}}; It is worth noting that we rule out repositories where Java is not the main language.
    (2) Repositories included in the Defects4j~\cite{defects4j} database, which is a dataset collecting reproducible bugs across multiple Java repositories.
    As a result, we collect $53$ repositories from GitHub and $17$ repositories from Defect4J, obtaining a total of $70$ candidate Java repositories.
    After careful manual selection and filtering, we narrowed down the list to $19$ open-source Java repositories: $10$ from source (1) and $9$ from source (2).
    \item \textbf{Issue instance crawling.}
    The issue instance crawling process was conducted in the following three steps:
    (1) We crawled all pull requests for the $19$ selected repositories.
    (2) We filtered these pull requests by retaining only those that were associated with at least one issue and involved changes to test files.
    (3) For each pull request that met our selection criteria, we further crawled its detailed information, including ``instance ID'', ``patch'', ``repository name'', ``base commit'', ``hints text'', ``creation date'', ``test patch'', ``issue statement'', ``environment setup commit'', and ``fail-to-pass''.
    In total, we crawled $1$,$979$ issue instances for $19$ repositories.    
    \item \textbf{Runtime environment determination.}
    We determine the runtime environment of each issue through code reading and trial runs.
    To be specific, we determine the build tool, JDK version and compilation commands used in the target issue.
    Given the code repository associated with the target issue, we carry out three steps: (1) We identify the build tool type (maven\footnote{\url{https://maven.apache.org}} or Gradle\footnote{\url{https://gradle.org}}) from the repository structure; 
    (2) We determine the JDK version used through reviewing the build configuration; 
    (3) We compile the repository to establish its compilation commands.
    In this phase, we also filter out issues where the associated repository fails to be compiled, e.g. the repositories depending on additional development environments like Android SDK.
    As a result, we primarily derive a collection of $308$ issue instances from the pool of $1$,$979$, which are verified to compile successfully under the determined runtime environment.
    \item \textbf{Fail-to-pass test extraction.}
    We extract fail-to-pass tests for each issue by comparing test results before and after applying the ground-truth patch.
    In detail, given an issue instance, we build two different containers to run the tests mentioned in the crawled test patch, respectively based on the code repository that has been patched with the fix and the code that hasn't.
    Then we parse the output test log to obtain the results of related tests, gathering those tests that fail before the issue fix but pass after it.
    Based on the test results, we also further conduct a filtering on issue instances.
    We only keep the issue instances which have at least one fail-to-pass test and no pass-to-fail test, deriving a set of 137 issue instances.
    \item \textbf{Questionnaire-based manual verification.} 
    To ensure the reliability of our benchmark in evaluating the LLMs' abilities in the issue resolving task, we conduct a comprehensive manual verification process.
    Following the recently published \texttt{SWE-bench-verified} annotation guidelines\footnote{\url{https://cdn.openai.com/introducing-swe-bench-verified/swe-b-annotation-instructions.pdf}}, we invite $10$ software developers experienced in Java, to screen the above $137$ issue instances. 
    The verification process involved answering the following three questions of each issue:
    (Q1) the clarity of the issue description, rated on a scale from $0$ to $3$, where lower scores indicate greater clarity;
    (Q2) the comprehensiveness of test coverage in evaluating potential solutions, also rated from $0$ to $3$, where lower scores reflect better coverage; 
    and (Q3) the presence of any major flaws that might necessitate exclusion from the dataset, with $0$ indicating the absence of such flaws and $1$ indicating their presence.
    Based on these annotations, we retained issues that satisfied the following criteria: ``\textit{(Q1 is 0 or Q1 is 1) and (Q2 is 0 or Q2 is 1) and (Q3 is 0)}''. This stringent filtering resulted in a final dataset of $91$ high-quality issue instances, covering $6$ repositories.
\end{enumerate}

\subsubsection{Troubleshooting}
During benchmark construction for \swebenchjava, we spot and fix a few issues which negatively affect the benchmark quality and evaluation efficiency.
Although \swebench has established a complete and easy-to-use pipeline for mining issues and automatically evaluating solutions of issue resolving in Python projects, we still encountered some difficulties when migrating to Java.
In this section, we will present the identified issues and discuss our solution for troubleshooting.

\paragraph{Base commit crawling errors in the original SWE-bench script.}
We found that the original SWE-bench script\footnote{\url{https://github.com/princeton-nlp/SWE-bench/tree/main/swebench/collect}} sometimes crawls the incorrect base commit for pull requests, mainly because it indiscriminately uses the previous commit without considering branch differences. We have fixed this bug using ``\texttt{git commit graph}'' to distinguish between different branches, making the script more reliable and ready for use.

\paragraph{Redundant downloads of repositories and dependencies.}
We found it a burden to repeatedly download repositories and dependencies when running evaluations on different issue instances.
To be specific, \swebench builds a separate docker container for each issue instance, where the corresponding repository will be pulled online to initialize the working directory; in addition, the dependencies will be installed to construct the runtime environment.
However, some of the issues share the same code repository and their dependencies may overlap, which means redundant downloads occur.
To reduce the redundancy of repository downloads, we pre-download all the selected repository locally all at once, and then replicate the downloaded repositories from local storage into containers.
We also plan to similarly maintain a local cache for dependencies and directly mount it to the built container to avoid repeated downloads.

\paragraph{Possible compilation broken during incremental compilation.}
When pre-compiling dependencies to improve caching, applying incremental patches can sometimes cause compilation failures.
Those failures may occur when the patches disrupt the project's build process, especially in projects with complex dependency structures.
To address this, we carefully analyzed the source code and project architecture, identifying potential conflicts. We then crafted specific test commands tailored to each project’s unique setup, ensuring that the build process remained stable despite the applied changes.
This thorough process helped us maintain consistent build integrity across different environments.

\subsection{Data Statistics}
\label{sec:data_stat}

\begin{figure}[!t]
    \centering
    \includegraphics[width=0.7\textwidth]{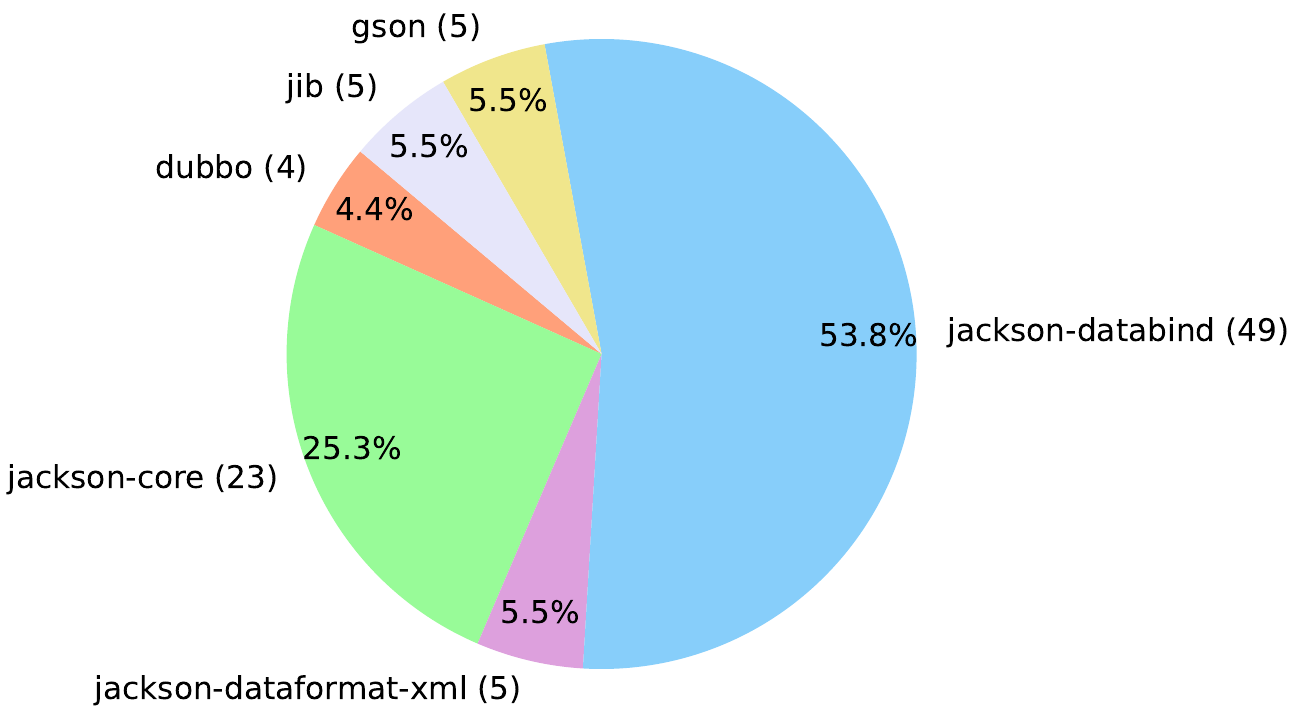}
    \caption{Distribution of \swebenchjava across $6$ GitHub repositories.}
    \label{fig:pie_repos_issues}
\end{figure}

As illustrated in Figure~\ref{fig:pie_repos_issues}, the \swebenchjava benchmark includes a total of $91$ issues across $6$ popular GitHub repositories. 
The distribution of issues varies among these repositories, with the highest concentration found in ``\texttt{fasterxml/jackson-databind}'' containing $49$ issues, while ``\texttt{apache/dubbo}'' has the fewest, with only $4$ issues.
These $6$ repositories span various domains, including data serialization (e.g., ``\texttt{fasterxml/jackson-core}'', ``\texttt{fasterxml/jackson-dataformat-xml}''), web services (e.g., ``\texttt{apache/dubbo}''), data formats (e.g., ``\texttt{google/gson}''), and container tools (e.g., ``\texttt{googlecontainertools/jib}''), demonstrating the dataset's broad coverage.
This diversity underscores the dataset's representativeness, providing a wide-ranging testbed for evaluating LLMs' performance in automated issue resolving within the Java ecosystem.

Table~\ref{tab:each_repo_details} provides a summary of key statistics for the repositories included in the \swebenchjava dataset, also highlighting their diversity and representativeness. 
The repositories vary widely in popularity, as indicated by their star counts, ranging from $0.56$K to $40.3$K, showcasing a broad selection of Java projects from different domains.
The repositories use either Maven or Gradle as build tools, with the majority being Maven projects.
The repository sizes also differ significantly, with repository sizes ranging from $57.4$K to $457.3$K lines of code, and file counts from $220$ to $4,387$.
This variety provides a comprehensive testbed for evaluating the difficulty of patch generation.
Besides, \swebenchjava often involves modifying multiple files (up to $105$ in the largest case), a substantial number of lines (up to $186$ lines), and numerous functions (up to $230$ functions).
Issues in these repositories vary in complexity, with character counts ranging from $1,241$ to $4,048$, and an average of $2,537$ characters, reflecting the detailed nature of the issue descriptions.

In summary, \swebenchjava's broad star distribution, varied build tools, diverse repositories, and issue complexities make it a practical and reliable benchmark for evaluating LLMs in the context of Java-specific automated issue resolving and patch generation.

\begin{table}[!t]
    \centering
    \caption{Summary statistics of \swebenchjava. ``Repository \#Files and \#Lines'' denote the total number of files and lines in the codebase, respectively. 
    ``Issue \#Chars'' indicates the average character count of issue descriptions. 
    ``Gold Patch \# Files, \# Lines and \# Func.'' represents the average number of files, lines, and functions modified per patch in the repository, while ``Test \#Lines'' denotes the total lines of code in the test cases.
    }
    \label{tab:each_repo_details}
    \scalebox{0.92}{
    \setlength{\tabcolsep}{4.3pt}
    \begin{tabular}{lrrrrrrrrr} 
\toprule
\multicolumn{1}{l|}{\multirow{2}{*}{\textbf{User/Repository}}} & \multicolumn{1}{c|}{\multirow{2}{*}{\textbf{ Star}}} & \multicolumn{1}{l|}{\multirow{2}{*}{\begin{tabular}[c]{@{}l@{}}\textbf{Build}\\\textbf{Tool}\end{tabular}}} & \multicolumn{2}{c|}{\textbf{Repository}}                   & \multicolumn{1}{c|}{\textbf{Issue}}    & \multicolumn{3}{c|}{\textbf{Gold Patch }}                                      & \multicolumn{1}{c}{\textbf{Test}}  \\
\multicolumn{1}{l|}{}                                          & \multicolumn{1}{c|}{}                                & \multicolumn{1}{l|}{}                                                                                       & \textbf{\# Files} & \multicolumn{1}{l|}{\textbf{\# Lines}} & \multicolumn{1}{c|}{\textbf{\# Chars}} & \textbf{\# Files} & \textbf{\# Lines} & \multicolumn{1}{l|}{\textbf{\# Func.}} & \textbf{\# Lines}                  \\ 
\hline\hline
\href{https://github.com/google/gson}{google/gson}                                                    &    23.2K                                                  &   Maven                                                                                                          & 296                  & 57.4K                                       &1,511                                        & 9                  & 89                 & 23                                       &  88                                  \\
\href{https://github.com/FasterXML/jackson-core}{FasterXML/jackson-core}                                         &  2.2K                                                    &  Maven                                                                                                           & 416                  &149.1K                                       & 1,241                                       &58                  & 125                  & 230                                       &  110                                  \\
\href{https://github.com/FasterXML/jackson-databind}{FasterXML/jackson-databind}                                     &  3.5K                                                    &  Maven                                                                                                           &1,259                   &303.2K                                        & 2,259                                       & 105                  & 75                  & 109                                       & 80                                   \\
\href{https://github.com/FasterXML/jackson-dataformat-xml}{FasterXML/jackson-dataformat-xml}                               &  0.56K                                                    &  Maven                                                                                                           & 220                  &  277.3K                                      & 3,794                                       &16                   &  186                 & 52                                       &   97                                 \\
\href{https://github.com/apache/dubbo}{apache/dubbo}                                                   & 40.3K                                                     & Maven                                                                                                            & 4,387                  &   457.3K                                     & 2,370                                       &12                   & 106               & 30                                       & 43                                  \\
\href{https://github.com/GoogleContainerTools/jib}{GoogleContainerTools/jib}                                       &13.6K                                                      &Gradle                                                                                                             & 1,195                  &102.7K                                        & 4,048                                       &13                   &  50                 &  16                                      & 53                                  \\
\hline
Mean                                                &--                                                      &--                                                                                                             & 2,537                  & 224.5K                                       &  2,537                                      & 36                  & 105                 &  77                                      &  79                                  \\
Max                                               &--                                                      &--                                                                                                            & 4,387                  & 457.3K                                       &  4,048                                      & 105                  & 186                  &  230                                      &  110                                  \\
\bottomrule
\end{tabular}
}
\end{table}


\section{Experiments}

\subsection{Experimental Setup}
\label{sec:exp_setup}

\subsubsection{Evaluation Metrics}

Following SWE-bench~\cite{swe-bench}, we adopt the Resolved Rate (\%) as our evaluation metric.
This metric indicates the proportion of issues in \swebenchjava that are successfully resolved.
An issue is considered resolved only if all given test cases pass.
This metric provides a precise measure of effectiveness in resolving real-world Java GitHub issues.

\subsubsection{Evaluation Approaches and Models}

To assess the reliability of \swebenchjava, we use SWE-agent~\cite{yang2024sweagent}, a classic approach that enhances LLM agents in software engineering tasks through a custom agent-computer interface (ACI). 
This interface empowers agents to autonomously create and edit code, navigate repositories, and execute tests.
With its proven superior performance on benchmarks like SWE-bench~\cite{swe-bench}, SWE-agent is well-suited for evaluating the robustness and practical value of our newly created \swebenchjava.

Note that we implemented SWE-agent rather hastily, without configuring the runtime environment for all Java issues in \swebenchjava.
This will affect the agent's ability to accurately reproduce issues, potentially leading to lower results (Section~\ref{sec:results}) compared to SWE-agent's normal performance.

Based on SWE-agent, we employ several popular and powerful models, including GPT-4o-2024-08-06\footnote{\url{https://openai.com/api/pricing}}, GPT-4o-mini-2024-07-18, DeepSeek-Coder-V2-0724\footnote{\url{https://platform.deepseek.com/api-docs/zh-cn/pricing}}, DeepSeek-V2-0628, and Doubao-pro-128k\footnote{\url{https://console.volcengine.com/ark/region:ark+cn-beijing/model/detail?Id=doubao-pro-128k}}.

\subsection{Results}
\label{sec:results}

Table \ref{tab:all_repo_details} demonstrates the varying capabilities of different models in solving \swebenchjava tasks. Compared to GPT and Doubao models, DeepSeek exhibits superior problem-solving abilities. The performance of GPT-4o significantly surpasses that of GPT-4o-mini, demonstrating the effectiveness of the model's comprehensive capabilities in problem-solving and highlighting the discriminative power of \swebenchjava in differentiating models. Overall, a greater amount of descriptive language in the issues leads to better problem-solving outcomes.

We observe that the more detailed the task description, the higher the requirement for the model's natural language understanding capability.
As observed in Table~\ref{tab:ratios_each_repo}, DeepSeek-V2 significantly outperforms DeepSeekCoder-V2 on the ``\texttt{GoogleContainerTools/jib}'' repository, which happens to have the most extensive natural language description among the six repositories (refer to Table\ref{tab:each_repo_details}). 
Similar results are also reflected in the ``\texttt{FasterXML/jackson-dataformat-xml}'' repository. Conversely, in the ``\texttt{FasterXML/jackson-core}'' repository, which has the least textual content, DeepSeekCoder-V2 performs notably better than DeepSeek-V2. This further corroborates the positive correlation between the level of detail in task descriptions and the required natural language understanding ability of the models.

The diverse performance of different models across \swebenchjava's various repositories underscores the diversity of this benchmark. 
Moreover, the significant performance disparities among models indicate that the dataset effectively distinguishes the capability differences between models, demonstrating good sensitivity and discrimination.

Furthermore, considering the scores achieved by the models, most tasks have not reached perfect or high scores, suggesting that \swebenchjava presents significant challenges to the work in related fields while also providing valuable guidance for future research. 
In summary, \swebenchjava reveals the limitations and strengths of models from multiple perspectives, holding great significance for advancing progress in relevant domains.

\begin{table}[t]
    \centering
    \caption{Resolved rate (\%) on \swebenchjava across various methods.}
    \label{tab:all_repo_details}
    \scalebox{1}{
    \begin{tabular}{lr} 
    \toprule
    \textbf{Methods}           & \textbf{Resolved Rate (\%)}  \\ 
    \hline\hline
    SWE-agent+GPT-4o-2024-08-06           &   $6.59$\% (6/91)                     \\
    SWE-agent+GPT-4o-mini-2024-07-18      &     $1.10$\% (1/91)                   \\
    SWE-agent+DeepSeekCoder-V2-0724 &      $7.69$\% (7/91)                  \\
    SWE-agent+DeepSeek-V2-0628 &       $9.89$\% (9/91)                 \\
    SWE-agent+Doubao-pro-128k           &   $1.10$\% (1/91)                     \\
    \bottomrule
    \end{tabular}
}
\end{table}

\begin{table}[t]
    \centering
    \caption{Proportion of issues resolved by each method across different repositories.}
    \label{tab:ratios_each_repo}
    \scalebox{0.98}{
    \begin{tabular}{lccccc} 
    \toprule
    \multirow{2}{*}{\textbf{User/Repository}} & \multicolumn{5}{c}{\textbf{SWE-agent }}                                                                                                         \\
                                              & \textbf{GPT-4o}            & \textbf{GPT-4o-mini}            & \textbf{DeepSeekCoder-V2}  & \textbf{DeepSeek-V2}       & \textbf{Doubao}             \\ 
    \hline\hline
    \href{https://github.com/google/gson}{google/gson}                               
        & $\filledcirclepercent{0}$ 
        & $\filledcirclepercent{0}$ 
        & $\filledcirclepercent{0}$ 
        & $\filledcirclepercent{0}$ 
        & $\filledcirclepercent{0}$  \\
    \href{https://github.com/FasterXML/jackson-core}{FasterXML/jackson-core}                    
        & $\filledcirclepercent{13.043478260869565}$ 
        & $\filledcirclepercent{0}$ 
        & $\filledcirclepercent{13.043478260869565}$
        & $\filledcirclepercent{0}$ 
        & $\filledcirclepercent{0}$  \\
    \href{https://github.com/FasterXML/jackson-databind}{FasterXML/jackson-databind}                
        & $\filledcirclepercent{4.081632653061225}$ 
        & $\filledcirclepercent{2.0408163265306123}$ 
        & $\filledcirclepercent{8.16326530612245}$ 
        & $\filledcirclepercent{12.244897959183673}$ 
        & $\filledcirclepercent{2.040816326530612}$  \\
    \href{https://github.com/FasterXML/jackson-dataformat-xml}{FasterXML/jackson-dataformat-xml}          
        & $\filledcirclepercent{0}$ 
        & $\filledcirclepercent{0}$ 
        & $\filledcirclepercent{0}$ 
        & $\filledcirclepercent{20}$ 
        & $\filledcirclepercent{0}$  \\
    \href{https://github.com/apache/dubbo}{apache/dubbo}                              
        & $\filledcirclepercent{0}$ 
        & $\filledcirclepercent{0}$ 
        & $\filledcirclepercent{0}$ 
        & $\filledcirclepercent{0}$ 
        & $\filledcirclepercent{0}$  \\
    \href{https://github.com/GoogleContainerTools/jib}{GoogleContainerTools/jib}                  
        & $\filledcirclepercent{20}$ 
        & $\filledcirclepercent{0}$ 
        & $\filledcirclepercent{0}$ 
        & $\filledcirclepercent{40}$
        & $\filledcirclepercent{0}$  \\
    \bottomrule
    \end{tabular}
    }
    \end{table}

\section{Related Works}
\label{sec:related_work}
The development of benchmarks for code generation has gained significant attention in recent years due to the increasing prominence of LLMs in programming tasks~\cite{nl2code,zhang2023survey,chen2023codet,pangucoder2,zan2024codes}.
These benchmarks serve as critical tools for evaluating the capabilities of LLMs in understanding, generating, and refining code. 
Early efforts in this domain focused on primarily evaluating models in monolingual settings~\cite{allamanis2013mining, raychev2016probabilistic, chen2021evaluating, austin2021program, wang2023execution, iyer2018mapping}.
For example, the HumanEval benchmark~\cite{chen2021evaluating} offers a range of Python programming problems of varying difficulty to evaluate the functional correctness of code generated by LLMs. 
Similarly, the MBPP benchmark~\cite{austin2021program} is widely used to assess LLMs' performance on basic Python programming problems.
As LLMs advanced, benchmarks became more sophisticated, evolving to better reflect real-world software engineering scenarios, such as library-oriented code generation, repository-level code completion, and issue resolving.
In the field of library-oriented code generation, several Python-specific benchmarks have emerged, including PandasEval~\cite{zan2022cert}, NumpyEval~\cite{zan2022cert}, and TorchDataEval~\cite{zan2022language}. These benchmarks assess the ability of LLMs to generate code that interacts effectively with popular libraries.
For repository-level code generation, early works like CoCoMIC~\cite{ding2024cocomic} and RepoEval~\cite{zhang2023repocoder,graphcoder} crafted datasets that require cross-file context to complete code, focusing exclusively on Python repositories. These benchmarks challenge LLMs to understand and complete code across multiple files within a repository.
In the field of repository-level issue resolving, SWE-bench~\cite{jimenez2024swebench} was created, including $2,294$ software engineering problems derived from GitHub issues and corresponding pull requests across $12$ Python repositories.
Resolving issues in \swebench often demands a deep understanding of the repository and the ability to coordinate changes across multiple functions, classes, and files simultaneously~\cite{coder,zhang2024autocoderover}.

In addition to monolingual benchmarks, there has been a growing interest in multilingual benchmarks to assess the performance of LLMs across different programming languages. 
For example, Multilingual-HumanEval~\cite{athiwaratkun2023multi} and HumanEval-X~\cite{zheng2023codegeex} extends HumanEval~\cite{codex} by providing equivalent tasks in multiple programming languages.
Similarly, MBXP~\cite{athiwaratkunmulti} extends the MBPP to include multilingual scenarios for more comprehensive evaluation.
MultiPL-E~\cite{cassano2023multipl} creates multilingual benchmarks based on HumanEval~\cite{codex} and MBPP~\cite{mbpp} in a translation-based way: it translates the two unit test-driven benchmarks to 18 additional programming languages.
This shift towards multilingualism reflects the global nature of software development and the need for LLMs to perform well across various programming contexts.
Moreover, recent work has emphasized the importance of benchmarking in real-world software engineering scenarios. 
The CoderEval~\cite{yu2024codereval} benchmark, for instance, is a context-aware benchmark for Java and Python that includes six levels of context dependency for code generation, such as class and file dependencies. 
Additionally, RepoBench~\cite{liurepobench} and CrossCodeEval~\cite{ding2024crosscodeeval} focus on more complex, real-world, multi-file programming tasks and thus serve as multilingual replacements of RepoEval~\cite{zhang2023repocoder}.

Despite these advancements, aligning benchmarks with real-world programming needs remains challenging. \swebench~\cite{jimenez2024swebench} provides a benchmark for evaluation in a realistic software engineering setting for Python repositories, which is monolingual. To address this, we extend \swebench to include Java, creating a multilingual benchmark to better evaluate LLMs' coding abilities in real-world scenarios.

\section{Conclusion and Future Works}
This paper presents \swebenchjava, an evaluation benchmark specifically designed for resolving issues in Java projects.
We detailed the construction process and conducted a comprehensive statistical analysis of the dataset.
Additionally, we have open-sourced the dataset, evaluation Docker environment, and leaderboard.
In the future, we plan to create benchmarks for more programming languages such as Go, Rust, C, and C++, while also continuing to improve the quality and coverage of the existing Java and Python datasets.

\section*{Acknowledgements}

We express our deepest gratitude to the creators of the SWE-bench~\cite{swe-bench} dataset, whose foundational work our project is built upon. 
We thank Lingzhi-zhiguang Co., Ltd for providing the computing resources used in this research.
This work was also supported by the National Key Research and Development Program of China under Grant No. 2022ZD0120201 - “Unified Representation and Knowledge Graph Construction for Science Popularization Resources”.

\bibliographystyle{unsrt}
\bibliography{references}

\begin{thebibliography}{10}

\bibitem{swe-bench}
Carlos~E Jimenez, John Yang, Alexander Wettig, Shunyu Yao, Kexin Pei, Ofir Press, and Karthik Narasimhan.
\newblock Swe-bench: Can language models resolve real-world github issues?
\newblock {\em arXiv preprint arXiv:2310.06770}, 2023.

\bibitem{yang2024sweagent}
John Yang, Carlos~E Jimenez, Alexander Wettig, Kilian Lieret, Shunyu Yao, Karthik Narasimhan, and Ofir Press.
\newblock {SWE-agent}: Agent-computer interfaces enable automated software engineering.
\newblock {\em arXiv preprint arXiv:2405.15793}, 2024.

\bibitem{nl2code}
Daoguang Zan, Bei Chen, Fengji Zhang, Dianjie Lu, Bingchao Wu, Bei Guan, Wang Yongji, and Jian-Guang Lou.
\newblock Large language models meet nl2code: A survey.
\newblock In {\em Proceedings of the 61st Annual Meeting of the Association for Computational Linguistics (Volume 1: Long Papers)}, pages 7443--7464, 2023.

\bibitem{zheng2023survey}
Zibin Zheng, Kaiwen Ning, Yanlin Wang, Jingwen Zhang, Dewu Zheng, Mingxi Ye, and Jiachi Chen.
\newblock A survey of large language models for code: Evolution, benchmarking, and future trends.
\newblock {\em arXiv preprint arXiv:2311.10372}, 2023.

\bibitem{zhang2023survey}
Ziyin Zhang, Chaoyu Chen, Bingchang Liu, Cong Liao, Zi~Gong, Hang Yu, Jianguo Li, and Rui Wang.
\newblock A survey on language models for code.
\newblock {\em arXiv preprint arXiv:2311.07989}, 2023.

\bibitem{codeStory}
SANDEEP~KUMAR PANI.
\newblock Our sota multi-agent coding framework.
\newblock {\em https://aide.dev/blog/sota-on-swe-bench-lite}, 2024.

\bibitem{defects4j}
Ren\'{e} Just, Darioush Jalali, and Michael~D. Ernst.
\newblock {Defects4J: a database of existing faults to enable controlled testing studies for Java programs}.
\newblock In {\em Proceedings of the 2014 International Symposium on Software Testing and Analysis}, ISSTA 2014, page 437–440, New York, NY, USA, 2014. Association for Computing Machinery.

\bibitem{huang2019empirical}
Yonghui Huang, Daniel~Alencar da~Costa, Feng Zhang, and Ying Zou.
\newblock An empirical study on the issue reports with questions raised during the issue resolving process.
\newblock {\em Empirical Software Engineering}, 24:718--750, 2019.

\bibitem{gpt4o}
OpenAI.
\newblock Hello gpt-4o.
\newblock 2024.
\newblock \url{https://openai.com/index/hello-gpt-4o}.

\bibitem{gpt4omini}
OpenAI.
\newblock Gpt-4o mini: advancing cost-efficient intelligence.
\newblock 2024.
\newblock \url{https://openai.com/index/gpt-4o-mini-advancing-cost-efficient-intelligence}.

\bibitem{deepseeklm}
DeepSeek-AI, :, Xiao Bi, Deli Chen, Guanting Chen, Shanhuang Chen, Damai Dai, Chengqi Deng, Honghui Ding, Kai Dong, Qiushi Du, Zhe Fu, Huazuo Gao, Kaige Gao, Wenjun Gao, Ruiqi Ge, Kang Guan, Daya Guo, Jianzhong Guo, Guangbo Hao, Zhewen Hao, Ying He, Wenjie Hu, Panpan Huang, Erhang Li, Guowei Li, Jiashi Li, Yao Li, Y.~K. Li, Wenfeng Liang, Fangyun Lin, A.~X. Liu, Bo~Liu, Wen Liu, Xiaodong Liu, Xin Liu, Yiyuan Liu, Haoyu Lu, Shanghao Lu, Fuli Luo, Shirong Ma, Xiaotao Nie, Tian Pei, Yishi Piao, Junjie Qiu, Hui Qu, Tongzheng Ren, Zehui Ren, Chong Ruan, Zhangli Sha, Zhihong Shao, Junxiao Song, Xuecheng Su, Jingxiang Sun, Yaofeng Sun, Minghui Tang, Bingxuan Wang, Peiyi Wang, Shiyu Wang, Yaohui Wang, Yongji Wang, Tong Wu, Y.~Wu, Xin Xie, Zhenda Xie, Ziwei Xie, Yiliang Xiong, Hanwei Xu, R.~X. Xu, Yanhong Xu, Dejian Yang, Yuxiang You, Shuiping Yu, Xingkai Yu, B.~Zhang, Haowei Zhang, Lecong Zhang, Liyue Zhang, Mingchuan Zhang, Minghua Zhang, Wentao Zhang, Yichao Zhang, Chenggang Zhao, Yao Zhao, Shangyan Zhou, Shunfeng
  Zhou, Qihao Zhu, and Yuheng Zou.
\newblock Deepseek llm: Scaling open-source language models with longtermism, 2024.

\bibitem{deepseekcoderv2}
DeepSeek-AI, Qihao Zhu, Daya Guo, Zhihong Shao, Dejian Yang, Peiyi Wang, Runxin Xu, Y.~Wu, Yukun Li, Huazuo Gao, Shirong Ma, Wangding Zeng, Xiao Bi, Zihui Gu, Hanwei Xu, Damai Dai, Kai Dong, Liyue Zhang, Yishi Piao, Zhibin Gou, Zhenda Xie, Zhewen Hao, Bingxuan Wang, Junxiao Song, Deli Chen, Xin Xie, Kang Guan, Yuxiang You, Aixin Liu, Qiushi Du, Wenjun Gao, Xuan Lu, Qinyu Chen, Yaohui Wang, Chengqi Deng, Jiashi Li, Chenggang Zhao, Chong Ruan, Fuli Luo, and Wenfeng Liang.
\newblock Deepseek-coder-v2: Breaking the barrier of closed-source models in code intelligence, 2024.

\bibitem{doubao}
Bytedance.
\newblock Doubao.
\newblock {\em https://console.volcengine.com/ark/region:ark+cn-beijing/model/detail?Id=doubao-pro-128k}, 2024.

\bibitem{chen2023codet}
Bei Chen, Fengji Zhang, Anh Nguyen, Daoguang Zan, Zeqi Lin, Jian-Guang Lou, and Weizhu Chen.
\newblock {CodeT: Code Generation with Generated Tests}.
\newblock In {\em The Eleventh International Conference on Learning Representations}, 2023.

\bibitem{pangucoder2}
Bo~Shen, Jiaxin Zhang, Taihong Chen, Daoguang Zan, Bing Geng, An~Fu, Muhan Zeng, Ailun Yu, Jichuan Ji, Jingyang Zhao, Yuenan Guo, and Qianxiang Wang.
\newblock {PanGu-Coder2: Boosting Large Language Models for Code with Ranking Feedback}, 2023.

\bibitem{zan2024codes}
Daoguang Zan, Ailun Yu, Wei Liu, Dong Chen, Bo~Shen, Wei Li, Yafen Yao, Yongshun Gong, Xiaolin Chen, Bei Guan, et~al.
\newblock {CodeS: Natural Language to Code Repository via Multi-Layer Sketch}.
\newblock {\em arXiv preprint arXiv:2403.16443}, 2024.

\bibitem{allamanis2013mining}
Miltiadis Allamanis and Charles Sutton.
\newblock Mining source code repositories at massive scale using language modeling.
\newblock In {\em 2013 10th working conference on mining software repositories (MSR)}, pages 207--216. IEEE, 2013.

\bibitem{raychev2016probabilistic}
Veselin Raychev, Pavol Bielik, and Martin Vechev.
\newblock Probabilistic model for code with decision trees.
\newblock {\em ACM SIGPLAN Notices}, 51(10):731--747, 2016.

\bibitem{chen2021evaluating}
Mark Chen, Jerry Tworek, Heewoo Jun, Qiming Yuan, Henrique Ponde De~Oliveira Pinto, Jared Kaplan, Harri Edwards, Yuri Burda, Nicholas Joseph, Greg Brockman, et~al.
\newblock Evaluating large language models trained on code.
\newblock {\em arXiv preprint arXiv:2107.03374}, 2021.

\bibitem{austin2021program}
Jacob Austin, Augustus Odena, Maxwell Nye, Maarten Bosma, Henryk Michalewski, David Dohan, Ellen Jiang, Carrie Cai, Michael Terry, Quoc Le, et~al.
\newblock Program synthesis with large language models.
\newblock {\em arXiv preprint arXiv:2108.07732}, 2021.

\bibitem{wang2023execution}
Zhiruo Wang, Shuyan Zhou, Daniel Fried, and Graham Neubig.
\newblock Execution-based evaluation for open-domain code generation.
\newblock In {\em Findings of the Association for Computational Linguistics: EMNLP 2023}, pages 1271--1290, 2023.

\bibitem{iyer2018mapping}
Srinivasan Iyer, Ioannis Konstas, Alvin Cheung, and Luke Zettlemoyer.
\newblock Mapping language to code in programmatic context.
\newblock In {\em Proceedings of the 2018 Conference on Empirical Methods in Natural Language Processing}, pages 1643--1652, 2018.

\bibitem{zan2022cert}
Daoguang Zan, Bei Chen, Dejian Yang, Zeqi Lin, Minsu Kim, Bei Guan, Yongji Wang, Weizhu Chen, and Jian{-}Guang Lou.
\newblock {CERT:} continual pre-training on sketches for library-oriented code generation.
\newblock In {\em Proceedings of the Thirty-First International Joint Conference on Artificial Intelligence, {IJCAI} 2022, Vienna, Austria, 23-29 July 2022}, pages 2369--2375, 2022.

\bibitem{zan2022language}
Daoguang Zan, Bei Chen, Zeqi Lin, Bei Guan, Wang Yongji, and Jian-Guang Lou.
\newblock When language model meets private library.
\newblock In {\em Findings of the Association for Computational Linguistics: EMNLP 2022}, pages 277--288, 2022.

\bibitem{ding2024cocomic}
Yangruibo Ding, Zijian Wang, Wasi~U Ahmad, Murali~Krishna Ramanathan, Ramesh Nallapati, Parminder Bhatia, Dan Roth, and Bing Xiang.
\newblock Cocomic: Code completion by jointly modeling in-file and cross-file context.
\newblock In {\em Proceedings of the 2024 Joint International Conference on Computational Linguistics, Language Resources and Evaluation (LREC-COLING 2024)}, pages 3433--3445, 2024.

\bibitem{zhang2023repocoder}
Fengji Zhang, Bei Chen, Yue Zhang, Jacky Keung, Jin Liu, Daoguang Zan, Yi~Mao, Jian-Guang Lou, and Weizhu Chen.
\newblock Repocoder: Repository-level code completion through iterative retrieval and generation.
\newblock In {\em Proceedings of the 2023 Conference on Empirical Methods in Natural Language Processing}, pages 2471--2484, 2023.

\bibitem{graphcoder}
Wei Liu, Ailun Yu, Daoguang Zan, Bo~Shen, Wei Zhang, Haiyan Zhao, Zhi Jin, and Qianxiang Wang.
\newblock {GraphCoder: Enhancing Repository-Level Code Completion via Code Context Graph-based Retrieval and Language Model}, 2024.

\bibitem{jimenez2024swebench}
Carlos~E. Jimenez, John Yang, Alexander Wettig, Shunyu Yao, Kexin Pei, Ofir Press, and Karthik~R. Narasimhan.
\newblock Swe-bench: Can language models resolve real-world github issues?
\newblock In {\em The Twelfth International Conference on Learning Representations, {ICLR} 2024, Vienna, Austria, May 7-11, 2024}, 2024.

\bibitem{coder}
Dong Chen, Shaoxin Lin, Muhan Zeng, Daoguang Zan, Jian-Gang Wang, Anton Cheshkov, Jun Sun, Hao Yu, Guoliang Dong, Artem Aliev, Jie Wang, Xiao Cheng, Guangtai Liang, Yuchi Ma, Pan Bian, Tao Xie, and Qianxiang Wang.
\newblock {CodeR: Issue Resolving with Multi-Agent and Task Graphs}, 2024.

\bibitem{zhang2024autocoderover}
Yuntong Zhang, Haifeng Ruan, Zhiyu Fan, and Abhik Roychoudhury.
\newblock Autocoderover: Autonomous program improvement, 2024.

\bibitem{athiwaratkun2023multi}
Ben Athiwaratkun, Sanjay~Krishna Gouda, Zijian Wang, Xiaopeng Li, Yuchen Tian, Ming Tan, Wasi~Uddin Ahmad, Shiqi Wang, Qing Sun, Mingyue Shang, et~al.
\newblock Multi-lingual evaluation of code generation models.
\newblock In {\em ICLR}, 2023.

\bibitem{zheng2023codegeex}
Qinkai Zheng, Xiao Xia, Xu~Zou, Yuxiao Dong, Shan Wang, Yufei Xue, Zihan Wang, Lei Shen, Andi Wang, Yang Li, Teng Su, Zhilin Yang, and Jie Tang.
\newblock Codegeex: A pre-trained model for code generation with multilingual benchmarking on humaneval-x.
\newblock In {\em Proceedings of the 29th ACM SIGKDD Conference on Knowledge Discovery and Data Mining}, pages 5673--5684, 2023.

\bibitem{codex}
Mark Chen, Jerry Tworek, Heewoo Jun, Qiming Yuan, Henrique~Ponde de~Oliveira~Pinto, Jared Kaplan, Harri Edwards, Yuri Burda, Nicholas Joseph, Greg Brockman, Alex Ray, Raul Puri, Gretchen Krueger, Michael Petrov, Heidy Khlaaf, Girish Sastry, Pamela Mishkin, Brooke Chan, Scott Gray, Nick Ryder, Mikhail Pavlov, Alethea Power, Lukasz Kaiser, Mohammad Bavarian, Clemens Winter, Philippe Tillet, Felipe~Petroski Such, Dave Cummings, Matthias Plappert, Fotios Chantzis, Elizabeth Barnes, Ariel Herbert-Voss, William~Hebgen Guss, Alex Nichol, Alex Paino, Nikolas Tezak, Jie Tang, Igor Babuschkin, Suchir Balaji, Shantanu Jain, William Saunders, Christopher Hesse, Andrew~N. Carr, Jan Leike, Josh Achiam, Vedant Misra, Evan Morikawa, Alec Radford, Matthew Knight, Miles Brundage, Mira Murati, Katie Mayer, Peter Welinder, Bob McGrew, Dario Amodei, Sam McCandlish, Ilya Sutskever, and Wojciech Zaremba.
\newblock Evaluating large language models trained on code, 2021.

\bibitem{athiwaratkunmulti}
Ben Athiwaratkun, Sanjay~Krishna Gouda, Zijian Wang, Xiaopeng Li, Yuchen Tian, Ming Tan, Wasi~Uddin Ahmad, Shiqi Wang, Qing Sun, Mingyue Shang, et~al.
\newblock Multi-lingual evaluation of code generation models.
\newblock In {\em The Eleventh International Conference on Learning Representations}.

\bibitem{cassano2023multipl}
Federico Cassano, John Gouwar, Daniel Nguyen, Sydney Nguyen, Luna Phipps-Costin, Donald Pinckney, Ming-Ho Yee, Yangtian Zi, Carolyn~Jane Anderson, Molly~Q Feldman, et~al.
\newblock Multipl-e: a scalable and polyglot approach to benchmarking neural code generation.
\newblock {\em IEEE Transactions on Software Engineering}, 49(7):3675--3691, 2023.

\bibitem{mbpp}
Jacob Austin, Augustus Odena, Maxwell Nye, Maarten Bosma, Henryk Michalewski, David Dohan, Ellen Jiang, Carrie Cai, Michael Terry, Quoc Le, and Charles Sutton.
\newblock Program synthesis with large language models, 2021.

\bibitem{yu2024codereval}
Hao Yu, Bo~Shen, Dezhi Ran, Jiaxin Zhang, Qi~Zhang, Yuchi Ma, Guangtai Liang, Ying Li, Qianxiang Wang, and Tao Xie.
\newblock Codereval: A benchmark of pragmatic code generation with generative pre-trained models.
\newblock In {\em Proceedings of the 46th IEEE/ACM International Conference on Software Engineering}, pages 1--12, 2024.

\bibitem{liurepobench}
Tianyang Liu, Canwen Xu, and Julian McAuley.
\newblock Repobench: Benchmarking repository-level code auto-completion systems.
\newblock In {\em The Twelfth International Conference on Learning Representations}.

\bibitem{ding2024crosscodeeval}
Yangruibo Ding, Zijian Wang, Wasi Ahmad, Hantian Ding, Ming Tan, Nihal Jain, Murali~Krishna Ramanathan, Ramesh Nallapati, Parminder Bhatia, Dan Roth, et~al.
\newblock Crosscodeeval: A diverse and multilingual benchmark for cross-file code completion.
\newblock {\em Advances in Neural Information Processing Systems}, 36, 2024.

\end{thebibliography}

\newpage

\appendix

\section*{Author List and Contributions}

\textbf{Daoguang Zan}: Project initiation, overall progress planning and management, experiment design, personnel coordination and task allocation, conceptual discussions, framework drafting, manuscript finalization, proofreading.

\textbf{Zhirong Huang}: Conceptual discussions, dataset collection and creation, data annotation, SWE-agent framework development, Section~\ref{sec:workflow_overview} (Benchmark Construction) writing, code writing and organization, server network debugging.

\textbf{Ailun Yu}: Conceptual discussions, dataset collection and creation, Docker evaluation environment configuration, data annotation, Section~\ref{sec:workflow_overview} (Benchmark Construction) framework and content writing, code writing and organization.

\textbf{Shaoxin Lin}: Conceptual discussions, data creation, data annotation.

\textbf{Yifan Shi}: 
Website development, including data visualization and continuous integration.

\textbf{Wei Liu}: Section~\ref{sec:related_work} (Related Works) draft writing, data annotation.

\textbf{Dong Chen}: Section~\ref{sec:introduction} (Introduction) draft writing, data annotation.

\textbf{Zongshuai Qi}: Section~\ref{sec:results} (Results) draft writing, data annotation, results analysis, proofreading.

\textbf{Hao Yu}: Section~\ref{sec:exp_setup} (Experimental Setup) draft writing, data annotation.

\textbf{Lei Yu}: Section~\ref{sec:data_stat} (Data Statistics) draft writing, data annotation, proofreading.

\textbf{Dezhi Ran}, \textbf{Muhan Zeng}: Data annotation, discussions.

\textbf{Pengjie Huang}: Initial idea discussions, computational resources provision.

\textbf{Bo Shen}, \textbf{Pan Bian}, \textbf{Guangtai Liang}, \textbf{Bei Guan}, \textbf{Tao Xie}, \textbf{Yongji Wang}: Guidance, discussions.

\textbf{Qianxiang Wang}: Project leadership.

\end{document}